%% file: main.tex
\documentclass[a4paper,11pt]{article}
\pdfoutput=1
\usepackage{jheppub} % for details on the use of the package, please see the JINST-author-manual
\arxivnumber{2310.02516
}

\usepackage{graphicx}
\usepackage{bm}
\usepackage{hyperref}
\usepackage{xcolor}
\usepackage{amsmath}
\usepackage{comment}
\usepackage{gensymb}
\usepackage{mathtools}

\let\Re\relax \let\Im\relax
\DeclareMathOperator{\Re}{Re}
\DeclareMathOperator{\Im}{Im}

\def\be{\begin{equation}}
\def\ee{\end{equation}}
\def\ba{\begin{eqnarray}}
\def\ea{\end{eqnarray}}

\usepackage{color}   %May be necessary if you want to color links
\definecolor{t}{rgb}{0.05,0.34,0.88}
\definecolor{bluh}{rgb}{0.0,0.1,0.9}
\usepackage{hyperref}
\hypersetup{
	colorlinks=true, %set true if you want colored links
	linktoc=all,     %set to all if you want both sections and subsections linked
	urlcolor=bluh,
	linkcolor=t,  %choose some color if you want links to stand out
	citecolor=bluh
}

\newcommand{\PT}{\mathcal{P}\mathcal{T}}
\newcommand{\CPT}{\mathcal{C}\mathcal{P}\mathcal{T}}
\newcommand{\ii}{\textrm{i}}
\newcommand{\mubar}{\bar{\mu}}
\newcommand{\lambdaR}{\lambda_{\textsc{r}}}
\newcommand{\gR}{g_{\textsc{r}}}

\newcommand{\pdrv}[2]{\frac{\partial#1}{\partial#2}}
\newcommand{\LambdaLP}{\Lambda_{\textsc{lp}}}
\newcommand{\calC}{\mathcal{C}}
\newcommand{\calD}{\mathcal{D}}

\newcommand{\upd}{\textrm{d}}
\newcommand{\Tc}{T_{\textsc{c}}}
\newcommand{\betaC}{\beta_{\textsc{c}}}
\newcommand{\Nmin}{N\textrm{--}1}
\newcommand{\minus}{\textrm{--}}

\newcommand{\tr}{\textrm{tr}}

\makeatletter
\newcommand{\superimpose}[3][\mathord]{#1{\mathpalette\superimpose@{{#2}{#3}}}}
\newcommand{\superimpose@}[2]{\superimpose@@{#1}#2}
\newcommand{\superimpose@@}[3]{%
  \ooalign{%
    \hfil$\m@th#1#2$\hfil\cr
    \hfil$\m@th#1#3$\hfil\cr
  }%
}
\makeatother

\newcommand{\sumintc}{\superimpose{\Sigma}{\textrm{\hspace{1pt}\scalebox{1.2}{$\int$}}}}

\title{Can negative bare couplings make sense? The \texorpdfstring{$\vec{\phi}^4$ theory at large $N$}{TEXT}}
\author{Ryan D. Weller}
\emailAdd{ryan.weller@colorado.edu}
\affiliation{Department of Physics, University of Colorado, Boulder, CO 80309, USA}

\abstract{Scalar $\lambda\phi^4$ theory in 3+1D, for 
a positive coupling constant $\lambda>0$, is known to have no interacting continuum limit, which is referred to as quantum triviality. However, it has been recently argued that the theory in 3+1D with an $N$-component scalar $\vec{\phi}$ and a $(\vec{\phi}\cdot\vec{\phi})^2=\vec{\phi}^4$ interaction term \textit{does} have an interacting continuum limit at large $N$. It has been suggested that this continuum limit has a negative (bare) coupling constant and exhibits asymptotic freedom, similar to the $\mathcal{P}\mathcal{T}$-symmetric $-g\phi^4$ field theory. In this paper I study the $\vec{\phi}^4$ theory in 3+1D at large $N$ with a negative coupling constant $-g<0$, and with the scalar field taking values in a $\mathcal{P}\mathcal{T}$-symmetric complex domain. The theory is non-trivial, has asymptotic freedom, and has a Landau pole in the IR, and I demonstrate that the thermal partition function matches that of the positive-coupling $\lambda>0$ theory when the Landau poles of the two theories (in the $\lambda>0$ case a pole in the UV) are identified with one another. The spirit of renormalization is that observables do not depend on the renormalization scale. Here we see even if the coupling is taken negative above the scale of the Landau pole, thermodynamic observables are unaffected. Thus the $\vec{\phi}^4$ theory at large $N$ appears to have a negative bare coupling constant; the coupling only becomes positive in the IR, which in the context of other $\PT$-symmetric and large-$N$ quantum field theories I argue is perfectly acceptable.}

\begin{document}
\maketitle
\flushbottom

\input intro.tex

\input calculation.tex

\input results.tex

\input conclusion.tex

\acknowledgments
    I would like to thank Paul Romatschke for many useful discussions. In addition, I would like to thank Seth Grable, Chun-Wei Su, and Sebastian Vazquez-Carson, as well as the heavy-ion theory group at CERN, including Urs Wiedemann and Jasmine Brewer, for conversations and comments on this subject. This work was supported by the Department of Energy, DOE award No. DE-SC0017905.

\bibliographystyle{JHEP}
\bibliography{bibliography.bib}

\end{document}

%% file: intro.tex
\section{Introduction}

Recently, Lagrangians with negative or even complex couplings have attracted interest in the context of renormalizable large-$N$ quantum field theories \cite{Berges:2023rqa,Romatschke:2022jqg,Grable2023,Romatschke:2022llf,Romatschke:2023sce}. As an example, $\vec{\phi}^4$ theory\footnote{Here, $\vec{\phi}^4$ is a shorthand (in rather abusive notation) for $(\vec{\phi}\cdot\vec{\phi})^2=(\vec{\phi}^2)^2$, which is the form of the interaction term in this theory.} with $N$-component scalars in 3+1D has been investigated by Romatschke non-perturbatively\footnote{Here, I mean non-perturbatively in the coupling.}
% The large-$N$ expansion is a kind of perturbative expansion. These results \textit{are} perturbative in large $N$, but not in the coupling.
in the large-$N$ limit. The theory has a Landau pole (even non-perturbatively in the coupling),
% (even non-perturbatively in the coupling!)
and it has been argued that the interacting continuum limit of the theory must have a negative bare coupling constant and exhibits asymptotic freedom \cite{Romatschke:2022jqg,Romatschke:2022llf, Romatschke:2023sce}, similarly to a proposal by Symanzik in the 70s for a $\phi^4$ theory with negative coupling that is asymptotically free \cite{osti_4024626,osti_4592657,Kleefeld:2005hf}. In fact the negative bare coupling in the $\vec{\phi}^4$ $\textrm{O}(N)$ model was pointed out as early as the 70s and 80s \cite{Kobayashi:1975ev,Bardeen:1983}. And a negative-coupling version of the theory was already understood to have asymptotic freedom \cite{Ogilvie:2008tu}.

To see if it is the case that there is a negative-coupling continuum limit of the theory, one should explicitly study the negative-coupling $\vec{\phi}^4$ theory in 3+1D, as will be done in this work. I will use the theory's path integral; however, a negative coupling constant in the Euclidean action for the $\vec{\phi}^4$ theory leads to an unbounded path integral over real-valued scalar fields $\vec{\phi}(x)\in\mathbb{R}^N$. Therefore, for the negative-coupling theory, the path integral must be made convergent by integrating over an appropriate half-dimensional subspace of all complex-valued scalar fields $\vec{\phi}(x)\in \mathbb{C}^N$.

As an additional complication, not every complex domain of path integration is ``appropriate''; that is, not every domain necessarily corresponds to a physical theory. For example, the path integral for the negative-coupling $\vec{\phi}^4$ theory in 1D converges on the domain $\vec{\phi}(\tau)=\vec{s}(\tau)e^{\ii\pi/4}$ for $\vec{s}(\tau)\in\mathbb{R}^N$, but this domain yields a complex-valued partition function\footnote{This corresponds to analytically continuing the eigenvalues of the positive-coupling Hamiltonian $E_n(\lambda)=E_n(1)\lambda^{1/3}$ to negative values of the coupling $\lambda<0$, i.e. rotating the $E_n(\lambda)$ by a complex phase factor $e^{\ii \pi /3}$.}. 
% (This corresponds to a complex-valued Hamiltonian spectrum when the Schr\"odinger eigenvalue problem is solved on this type of domain.) 
Certain complex domains of path integration do, nonetheless, correspond to a physical partition function. Such domains appear, for example, in $\PT$-symmetric quantum theories (see e.g. the construction of the path integral in \cite{Bender:2006wt}). 

$\PT$-symmetric theories have non-Hermitian Hamiltonians with negative or complex couplings and yet can exhibit spectra that are real and bounded from below, as was first discovered by Bender and Boettcher \cite{Bender:1998ke}. With the additional construction of an appropriate $\mathcal{C}$ operator and a $\CPT$ inner product of states \cite{Bender:2002vv}, probabilities in $\PT$-symmetric theories can be defined and calculated. Here it should be noted that $\phi^4$ field theory (for $N=1$) with a negative coupling constant can be given meaning as a $\PT$-symmetric theory and in 3+1D it is asymptotically free \cite{Bender:1999ek,Ai:2022olh}. 

It is interesting to note that, similar to the $N$-component $\vec{\phi}^4$ theory in 3+1D, the Lee model \cite{PhysRev.95.1329} from the 1950s also exhibits a divergence in the coupling at some critical scale, above which the squared coupling $g^2$ becomes negative. In this case the Lee model can be interpreted as a quantum field theory with a non-Hermitian Hamiltonian and a negative squared coupling constant, as pointed out by Kleefeld \cite{Kleefeld:2004jb}. In fact, the Lee model can be interpreted as a $\PT$-symmetric theory, and the negative-norm ghost states that appear from the renormalization can be reinterpreted as physical (positive-norm) states when the $\CPT$ inner product is introduced \cite{Bender:2004sv,Bender:2007nj}. Thus, Hermiticity is not required for a physical theory, as has been long understood by the $\PT$ symmetry community. For example, non-Hermitian Hamiltonians appear in lossy systems. And Hermiticity can be replaced, for example by (unbroken) $\PT$ symmetry, or in general an antilinear symmetry \cite{Mannheim:2015hto}, as a requirement for a theory with a positive-definite inner product and unitarity. Moreover, the example of the Lee model suggests that a theory can have a negative bare coupling constant. I note that non-Hermitian $\PT$-symmetric systems have been realized experimentally, such as in systems with coupled gain and loss, and in optical analogs (e.g. \cite{Li2019PTFloquetUltracoldAtoms,Wu2019SingleSpinPTBreaking,Naghiloo2019TomographyEPQubit,Ruter2010ObservationPTOptics,Peng2014PTWGMicrocavities,Feng2014SingleModePTLaser}).

On the path integral level, a $\PT$-symmetric theory simply corresponds to integrating over any complex domain that asymptotically terminates within certain regions called Stokes wedges\footnote{See equation (61) in Bender \textit{et al.} \cite{Bender:2006wt} and the surrounding discussion. The path integral of the $\PT$-symmetric theory is simply defined as an integral over an appropriate complex domain $\calC_{\PT}$. The $\calC$ operator does not show up explicitly in the path integral picture. See also \cite{Jones:2006sj} for this last point. Although there may be a $\mathcal{C}\mathcal{P}\mathcal{T}$ inner product which is non-trivial to calculate, it doesn't enter into a Lagrangian description of a $\mathcal{P}\mathcal{T}$-symmetric theory.}. For the $-g\phi^4$ theory these lie at angles in the lower half of the complex plane \cite{Bender:1998ke}. Therefore, even with a negative coupling constant, there exist domains of path integration that give physical theories. Because such domains do exist, a negative-coupling theory can have meaning; that is, the theory can be predictive. Making sense of the negative-coupling theory simply amounts to finding such a domain.

An early analysis of $\mathcal{PT}$-symmetric negatively-coupled $\textrm{O}(N)$ models at large $N$ was carried out by Ogilvie and Meisinger \cite{Ogilvie:2008tu}, who derived the large-$N$ effective description using a constraint-field formulation, with which the present results are consistent. In that work, the authors emphasized that determining stability in the $\mathcal{PT}$-symmetric $\textrm{O}(N)$ field theory requires a careful understanding of the domains used in functional path integration (which the following work attempts to address), and they asserted that renormalization of the quartic coupling in four dimensions leads to an asymptotically free theory in the large-$N$ limit. However, the analysis remained schematic in several respects: no explicit convergent domain of path integration for the scalar fields was specified, and the renormalization was not carried out in detail. In addition, no comparison was made between renormalized thermodynamic quantities and those of the corresponding positive-coupling (supposedly Hermitian) theory. The present work extends their work in this way.

In \cite{Romatschke:2022jqg,Romatschke:2022llf} that studied the $N$-component $\vec{\phi}^4$ model,  the partition function for the negative-coupling theory was not calculated directly. Rather, it was calculated assuming the unproven recent conjecture by Ai, Bender, and Sarkar in \cite{Ai:2022csx} for the $\PT$-symmetric $-g\phi^4$ theory. This conjecture holds that the logarithm of the thermal partition function $Z_{\PT}(g,\beta=1/T)$ at negative coupling $-g<0$ is given from the partition function $Z_+(\lambda,\beta)$ calculated at positive coupling $\lambda$ as
\begin{equation}
\ln Z_{\PT}(g,\beta) = \frac{1}{2}\ln Z_+(\lambda\to-g+\ii 0^+,\beta) + \frac{1}{2}\ln Z_+(\lambda\to-g-\ii 0^+,\beta).
\end{equation}
I do not expect the reader to know of this conjecture; the main point to take away is that is that this present paper tries to derive the same result without relying upon any conjectures. As an additional point, the earlier \cite{Ogilvie:2008tu} calculated up to equation \eqref{eq:pressure-per-component-simple} in this work, yet made no connection to the positive-coupling theory after renormalization.
Therefore, there has not yet been an explicit demonstration that the negative-coupling theory (path-integrated on some appropriate domain) really is continuum limit of the positive-coupling theory. Such a demonstration is the aim of this paper.

That is to say, in this paper I show that there exists at least some complex domain of path integration for which the negative-coupling $\vec{\phi}^4$ theory at large $N$ has the same thermal partition function as the positive-coupling theory. This domain respects the $\PT$ symmetry $\vec{\phi}\to-\vec{\phi}$, $\ii\to-\ii$ 
% \footnote{The $\PT$ symmetry can also include actual spacetime reflections $\mathbf{x}\to-\mathbf{x}$, $t\to-t$, without affecting the results of this work.}
(although it is not the usual domain one might consider for the $\PT$-symmetric theory, which I will define). In the standard picture of renormalization, observables are unchanged by altering the bare coupling and renormalization scale along renormalization group flows. Here I will show that thermodynamic observables are unaffected by taking the renormalization scale to be large and the bare coupling to be negative. This lack of change in physical observables under a change of the bare coupling suggests that in some sense the negative-coupling theory ``flows'' to the positive-coupling $\textrm{O}(N)$ model in the IR, hence the calculation \textit{suggests} that the model with negative coupling is the continuum limit of the $\vec{\phi}^4$ theory at large $N$. However, more work should be done to establish if this is truly the case. 

This paper is organized as follows. First, I review the result at large $N$ for the partition function for the positive-coupling $\vec{\phi}^4$ theory (i.e. the standard $\textrm{O}(N)$ model) in 3+1D. Then, I introduce the negative-coupling theory. I define a domain on which the partition function of the $-g\vec{\phi}^4/N$ theory can be calculated at large $N$. Next, I renormalize the coupling $g$, and demonstrate that the partition function of the negative-coupling theory on this domain is equivalent to that of the positive-coupling theory after renormalization. We will see that the negative-coupling theory is asymptotically free and has no Landau pole in the UV, giving evidence that the $\vec{\phi}^4$ theory at large $N$ really is non-trivial and has an interacting continuum limit, as was argued by Romatschke \cite{Romatschke:2023sce}. Lastly, I'll conclude with remarks about future directions of this work.

%% file: calculation.tex
\section{Calculation}

In the large-$N$ expansion, the thermal partition function of the theory can be written as
\begin{equation}
    Z\propto e^{N \beta V p(\beta)},
    \label{eq:large-N-Z}
\end{equation}
where $V\to\infty$ is the volume of space and $p(\beta)$ is the pressure per component as a function of inverse temperature $\beta=1/T$. At leading order in large $N$, $p(\beta)$ will not depend on $N$. 
% At higher orders in the large-$N$ expansion, there will be corrections of order $\mathcal{O}(1/N^k)$ to $p(\beta)$, with $k\in\mathbb{Z}_+$ a positive integer. 

In this section I first review the leading-order result for the partition function $Z_+$ of the positive-coupling theory, and then I calculate the partition function $Z_-$ for the negative-coupling theory defined on some complex domain $\calC_-$ of path integration. 

\subsection{Review of the positive-coupling theory}

The thermal partition function of the positive-coupling theory is formally given by
\begin{equation}
    Z_+\propto\int \calD^{N}\vec{\phi} \,e^{-\int S_+[\vec{\phi}]}
\end{equation}
where the path integral is over real-valued fields $\vec{\phi}(x)\in\mathbb{R}^N$ with periodic boundary conditions in Euclidean time $\tau$: $\vec{\phi}(\tau+\beta,\mathbf{x})=\vec{\phi}(\tau,\mathbf{x})$. The Euclidean action $S_+[\vec{\phi}]$ is 
\begin{equation*}
    S_+[\vec{\phi}]=\int_{\beta,V} \upd^4x \,\bigg( \frac{1}{2} (\partial_\mu\vec{\phi})^2+\frac{\lambda}{N}\vec{\phi}^4 \bigg),
\end{equation*}
where $\int_{\beta,V}\upd^4x=\int_0^{\beta}\upd\tau\int_V
    \upd^3\mathbf{x}$. In this work I only consider the massless case, but it is easy to generalize to the case with a mass term $M\vec{\phi}^2$ as done in \cite{Romatschke:2022llf,Romatschke:2023sce}.

For this  theory, the expression for $p(\beta)$ in \eqref{eq:large-N-Z} has already been calculated and can be found for example in \cite{Romatschke:2022llf}. In dimensional regularization in (3--$2\epsilon$)+1D, the expression is
\begin{equation}
\begin{split}
    p(\beta)=\frac{m^4}{16\lambda}+\frac{m^4}{64\pi^2}\bigg(\frac{1}{\epsilon} + \ln\bigg(\frac{\mubar^2}{m^2}\bigg)+\frac{3}{2}\bigg) 
    + \frac{m^2}{2\pi^2\beta^2}\sum_{n=1}^{\infty} \frac{K_2(n\beta m)}{n^2}.
\end{split}
\label{eq:p-per-component-pos-coup}
\end{equation}
Here, $\mubar$ is the $\overline{\textrm{MS}}$ scale and $K_2(x)$ is a modified Bessel function of the second kind. $m$ is a gap parameter that is fixed by solving the gap equation $\partial p(\beta) / \partial m^2 = 0$. 
% More specifically, the Lefschetz anti-thimble passing through the gap solution $m^2=\ii\zeta_0$ should intersect the real line $\zeta_0\in\mathbb{R}$ in order for the solution to be relevant (i.e. to contribute its pressure to the partition function). 
Only the 
% relevant
solution to the gap equation which gives the larger pressure contributes in the large-volume limit.

\begin{figure}
    \centering
    \includegraphics[width=75mm]{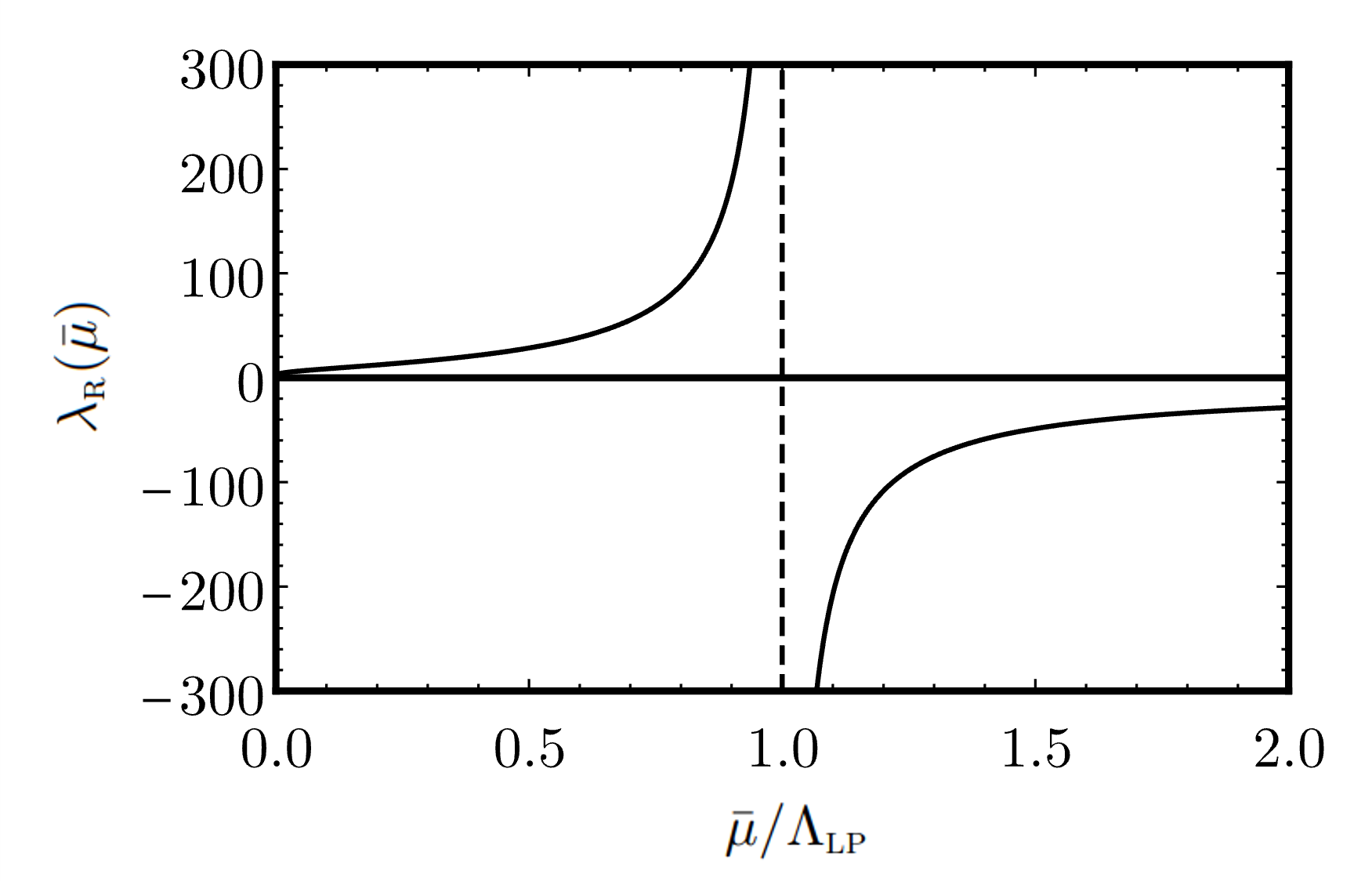}
    \caption{A plot of the renormalized running coupling $\lambdaR(\mubar)$ from equation \eqref{eq:running-coupling} in the $\lambda\vec{\phi}^4/N$ theory as a function of $\overline{\textrm{MS}}$ scale $\mubar$. One sees that at the scale $\mubar=\LambdaLP$ of the Landau pole, the coupling diverges, and above that scale, the coupling becomes negative.}
    \label{fig:Landau-pole}
\end{figure}

The expression \eqref{eq:p-per-component-pos-coup} for $p(\beta)$ has a $1/\epsilon$ divergence from the dimensional regularization which can be absorbed into the renormalized coupling $\lambdaR(\mubar)$ by defining
\begin{equation}
    \frac{1}{\lambdaR(\mubar)}=\frac{1}{\lambda(\mubar)}+\frac{1}{4\pi^2\epsilon}.
    \label{eq:ren-coupling}
\end{equation}
The renormalized coupling depends on $\mubar$ such that the pressure per component $p(\beta)$ does not depend on $\mubar$. This requirement, $\partial p(\beta) / \partial \mubar = 0$, which gives the beta function, fixes the running coupling to be
\begin{equation}
    \lambdaR(\mubar)=\frac{4\pi^2}{\ln(\LambdaLP^2/\mubar^2)},
    \label{eq:running-coupling}
\end{equation}
with $\LambdaLP$ a scale that appears from the constant in the integration of the beta function. $\LambdaLP$ is determined for example by fixing the value $\lambda_0$ of the coupling $\lambdaR(\Lambda_0)=\lambda_0$ at some UV scale $\Lambda_0$. The running of the coupling is visualized in figure \ref{fig:Landau-pole}.

One immediate objection can be made. If the coupling has some non-zero, positive value at some finite scale $\mubar$, then the scale $\LambdaLP>\mubar$ will also be finite. Above the scale $\LambdaLP$, the coupling in \eqref{eq:running-coupling} becomes negative. Therefore, there is no way to take the 
UV scale $\Lambda_0$ to infinity and keep the UV coupling $\lambda_0$ positive. Since a positive coupling was assumed in order to make the calculation \eqref{eq:p-per-component-pos-coup} of $p(\beta)$ and of the partition function $Z_+$, this is problematic\footnote{However, I am being a little heuristic. It is really the coupling $\lambda$, and not $\lambdaR$, that was assumed to be positive here. In dimensional regularization in (3--2$\epsilon$)+1D this means that $\lambda(\mubar)$ becomes negative at a scale $\mubar=\LambdaLP\, e^{-1/2\epsilon}$. The fact remains that the coupling diverges and becomes negative at some scale. In cutoff regularization with cutoff $\Lambda$, there is a similar scale $\Lambda=\LambdaLP$ above which the coupling $\lambda(\Lambda)$ becomes negative \cite{Romatschke:2023sce}, and $\lambda(\Lambda)$ is identical to $\lambdaR(\mubar)$ in \eqref{eq:running-coupling} if $\mubar$ is identified with $\Lambda$.}. The interacting theory appears to have no continuum limit $\Lambda_0\to\infty$, and if one tries to fix this problem by taking $\LambdaLP$ to infinity, the coupling at any finite scale $\mubar<\infty$ will be zero, such that the theory is trivial (non-interacting). This is a feature of scalar $\phi^4$ field theory, which is known to be quantum trivial in 3+1D for $N=1,2$ components \cite{FROHLICH1982281,Aizenman:2019yuo}. 

As an additional note, the coupling $\lambdaR(\mubar)$ will diverge at the scale $\mubar=\LambdaLP$. Therefore the $\vec{\phi}^4$ theory at large $N$ has a Landau pole even non-perturbatively (as opposed to only perturbatively). 
% This disproves the wives' tale that a Landau pole simply signals the breakdown of the perturbative approximation and is an artefact of doing loop perturbation theory in the coupling. Here, in eq. \eqref{eq:running-coupling}, there is still a Landau pole, even though the calculation has been done entirely non-perturbatively in the coupling.

Despite these potential issues, assuming that $\mubar<\LambdaLP$ and that the $\vec{\phi}^4$ theory is simply an effective (cut-off) theory, the calculation of the pressure $p(\beta)$ in \eqref{eq:p-per-component-pos-coup} can proceed, and has been done by Romatschke, and the result plotted, in \cite{Romatschke:2022jqg,Romatschke:2022llf}. The renormalized pressure is given by
\begin{equation}
\begin{split}
           p(\beta)=\frac{m^4}{64\pi^2}\bigg(\ln\bigg(\frac{\LambdaLP^2}{m^2}\bigg)+\frac{3}{2}\bigg)
           + \frac{m^2}{2\pi^2\beta^2}\sum_{n=1}^{\infty} \frac{K_2(n\beta m)}{n^2}.
\end{split}
    \label{eq:p-per-component-pos-coup-renormed}
\end{equation}
At low temperatures $T\leq\Tc\approx0.616\,\LambdaLP$, one finds two solutions $m$ to the gap equation $\partial p(\beta)/\partial m^2=0$. The dominant solution has $m^2=e\LambdaLP^2$ at $T=0$ and determines the pressure $p(\beta)$ via \eqref{eq:p-per-component-pos-coup-renormed}. At higher temperatures $T>\Tc$ there are also two solutions to the gap equation, but now they are a complex-conjugate pair of solutions $m_+,m_-=m_+^*$ corresponding to complex-conjugate pressures $p_+(\beta),p_-(\beta)=p_+^*(\beta)$. Putting the partition function $Z_+$ into the large-$N$ form of \eqref{eq:large-N-Z} leaves one with
\begin{equation}
\begin{split}
       Z_+ \propto e^{N \beta V \Re p_+(\beta)+\ln \cos\big(N\beta V \Im p_+(\beta)\big)}
       \approx e^{N \beta V \Re p_+(\beta)},\,\,\,\,\beta<\betaC,
\end{split}
\label{eq:large-N-Z-2}
\end{equation}
which follows formally at large $N$ from only keeping the part of $\ln Z_+$ that scales like $N$ \footnote{Note that the argument for setting $p(\beta)=\Re p_+(\beta)$ in \cite{Romatschke:2022jqg,Romatschke:2022llf} was based on the unproven conjecture in \cite{Ai:2022csx} for the $\PT$-symmetric $-g\phi^4$ theory. Here instead I've alternatively based it on the formal expansion of $\ln Z$ in large $N$. Later, in subsection \ref{subsec:high-temps}, I discuss the consequences of not neglecting the imaginary parts of both saddles, and how it suggests $\PT$ symmetry breaking at high $T$.}. 
% \footnote{
% There is some large-$N$ magic happening here.
% because we are doing an analytic expansion in powers of $1/N$. Non-analytic terms are not ``known about'' by this expansion. It is probably true that the large-$N$ expansion gives an asymptotic (divergent) series. But
% $\ln\cos (N\cdots)$ is formally ``small'' because it spends most of its time close to $0$ and to $\ii \pi$ as $N\to \infty$.}$^,$
One finds that $p(\beta<\betaC)=\Re p_+(\beta)$ is physically well-behaved (e.g. increases with temperature) and $p(\beta)$ has a continuous first derivative at $\beta=\betaC$. 

Interestingly enough, the pressure per component $p(\beta)$ of the $\lambda\vec{\phi}^4/N$ theory, calculated in this way, asymptotes toward the Stefan--Boltzmann limit for a single free boson at high temperatures $\beta\to0$ \cite{Romatschke:2022jqg}, suggesting that the $\vec{\phi}^4$ theory at large $N$ is asymptotically free. This should not be true based on the argument that the Landau pole in \eqref{eq:running-coupling} prevented the positive-coupling theory from having an interacting continuum limit. Moreover, the beta function $\partial \lambdaR / \partial \ln \mubar$ has the same sign as the coupling assuming $\mubar<\LambdaLP$. Asymptotically free theories, on the other hand, are interacting, do have a continuum limit, and have a beta function whose sign is opposite that of the coupling as $\mubar\to\infty$.

The solution to this this apparent contradiction, and to the problem of triviality, is to allow the bare coupling to be negative. Non-triviality and a negative bare coupling were argued in \cite{Romatschke:2023sce}. But if the bare coupling is negative, the calculation of $p(\beta)$ must be  carefully done in some other way that assumes negative coupling from the start. 

That is one new result and the next part of this work.

\subsection{Defining the negative-coupling theory}

The thermal partition function for the negative-coupling theory is given by
\begin{equation}
    Z_-\propto\int_{\calC_-} \calD^{N}\vec{\phi} \,e^{-\int S_-[\vec{\phi}]}
    \label{eq:part-func-neg-coup}
\end{equation}
where the fields still have periodic boundary conditions in Euclidean time
% \footnote{One \textit{might} expect a $\PT$-symmetric quantum theory to have $\calC\PT$-twisted boundary conditions, which would look like $\calC\vec{\phi}(0,\mathbf{x})=-\vec{\phi}^*(\beta,\mathbf{x})$, but I have not seen mention of this in the literature. Such boundary conditions are not included, for example, in the zero-temperature formulation of the path integral in section III of \cite{Ai:2022csx} which studied the $-g\phi^4$ field theory.}
but now the path integral is over complexified fields that live on the domain $\calC_-$
(which will be specified shortly). The Euclidean action $S_-[\vec{\phi}]$ is 
\begin{equation}
    S_-[\vec{\phi}]=\int_{\beta,V}
    \upd^4x \,\bigg( \frac{1}{2} (\partial_\mu\vec{\phi})^2-\frac{g}{N}\vec{\phi}^4 \bigg),
    \label{eq:action-neg-coup}
\end{equation}
where $-g<0$ is now the coupling. 

A choice 
for the domain $\calC_-$ 
that one might consider for the $\PT$-symmetric $-g\vec{\phi}^4/N$ theory is the domain $\calC_{\PT}$ parametrized by $N$ real-valued fields $\chi(x)\in\mathbb{R}$, $\vec{\eta}(x)\in\mathbb{R}^{\Nmin}$ as
\begin{equation}
    \begin{split}
        \vec{\phi}(x)&\cdot\hat{e}=
        \chi(x) f\big(\chi(x)\big)\\
        \vec{\phi}(x)-&\big(\vec{\phi}(x)\cdot\hat{e}\big)\hat{e}=\vec{\eta}(x) f\big(\chi(x)\big),
    \end{split}
    \label{eq:domain-PT}
\end{equation}
where $\hat{e}$ is some unit vector in $\mathbb{R}^N$ and the function $f$ is defined as
\begin{equation}
f(\chi)\equiv  \theta\big(\chi)e^{-\ii\pi/4}+\theta(\minus\chi)e^{\ii\pi/4}, 
\label{eq:f}
\end{equation}
where $\theta$ is the Heaviside step function. This forces one component $\vec{\phi}(x)\cdot\hat{e}$ of the field $\vec{\phi}$ to live on the union of two half-lines at angles $-\pi/4$, $-3\pi/4$ in the lower half of the complex plane at every point $x$ in Euclidean spacetime. If every component lived independently on the same union of half-lines, the path integral would be unbounded, as was pointed out in \cite{Lawrence:2023woz}. This domain respects the $\PT$ symmetry $\vec{\phi}\to-\vec{\phi}$, $\ii\to-\ii$, which is also a symmetry of the action $S_-[\vec{\phi}]$, and any deformation of this domain with the same Lefschetz thimble decomposition (e.g.
terminating within the associated Stoke's wedges)
gives the same partition function $Z_-=Z_{\PT}$.

\begin{figure}
    \centering
    \includegraphics[width=75mm]{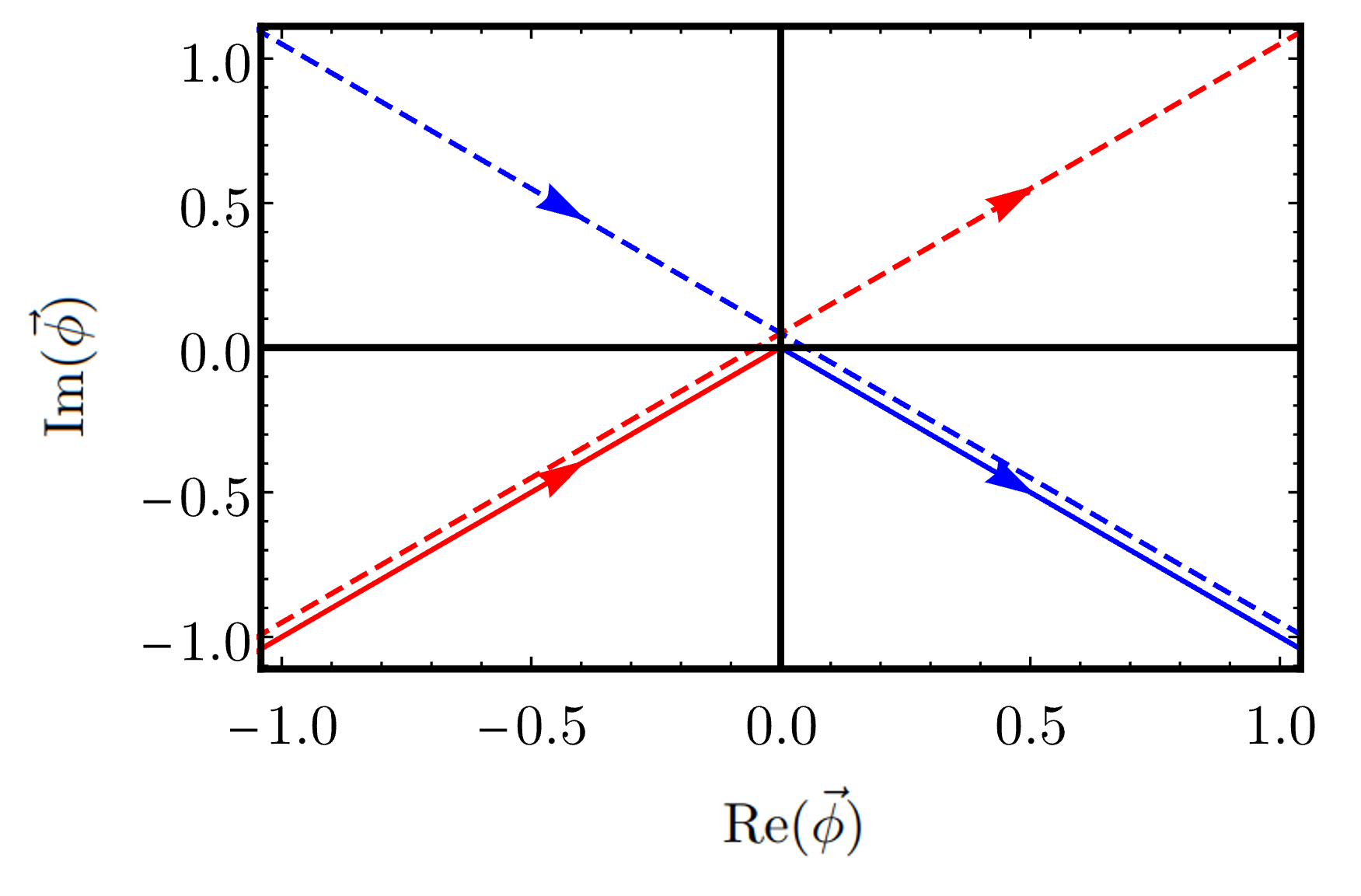}
    \caption{A visualization of the $\PT$-symmetric domains of path integration $\calC_{\PT}$ in \eqref{eq:domain-PT} and $\calC_-$ in $\eqref{eq:domain-neg-coup}$. The axes are expressed in some unit of mass (it does not matter which). For the case of $\calC_{\PT}$, $\vec{\phi}(x)\cdot\hat{e}$ lives on either the solid red line or the solid blue line, and every component of $\vec{\phi}(x)-\big(\vec{\phi}(x)\cdot\hat{e}\big)\hat{e}$ lives on the corresponding dashed line of the same color (red or blue). For the case of $\calC_-$, the zero mode $\vec{\phi}_0\cdot\hat{e}$ lives on either the solid red or the solid blue line, while the non-zero modes $\vec{\phi}'(x)\cdot\hat{e}$ and every component of $\vec{\phi}(x)-\big(\vec{\phi}(x)\cdot\hat{e}\big)\hat{e}$ live on the corresponding dashed line of the same color.}
    \label{fig:PT-domains}
\end{figure}

The path integral on the domain $\calC_{\PT}$ in \eqref{eq:domain-PT}, however, is difficult to solve analytically using standard large-$N$ techniques, because there are an infinite number of sharp kinks in the domain $\calC_{\PT}$, with one kink at $\chi(x)=0$ for every point $x$ in spacetime. These kinks can be managed straightforwardly in a lattice calculation (and the $N=1$ lattice calculation in 3+1D was done in \cite{Romatschke:2023sce} and in 1D in \cite{Lawrence:2023woz}), but on the lattice the path integral \eqref{eq:part-func-neg-coup} with $\calC_-=\calC_{\PT}$ from \eqref{eq:domain-PT} has a sign problem, so the calculation is difficult to scale to $N$ components. 

One could consider a domain $\calC_{\PT}$ without kinks by 
generalizing the hyperbola $\phi(x)=-2\ii\sqrt{1+\ii\chi(x)}$ for $\chi(x)\in\mathbb{R}$ in \cite{Bender:2006wt,Jones:2006qs} to $N$-component scalar fields, but it is unclear how to do this. One such generalization to $N$ components for the 1D theory was done in \cite{Bender:2013mpa} by letting the radial coordinate live on this complex hyperbola, i.e. $\vec{\phi}^{\,2}(x)=-4\big(1+\ii \chi(x)\big)$ for $\chi(x)\in\mathbb{R}$, but this generalization gives an unbounded Hamiltonian spectrum and thus an unbounded partition function in 1D once angular momentum is considered\footnote{The interested reader can check this by including the angular momentum quantum number $\ell$ in the radial Schr\"odinger equation and finding the Hermitian Hamiltonian which is equivalent to the $\PT$-symmetric non-Hermitian Hamiltonian along the lines of \cite{Jones:2006qs,Bender:2006wt,Bender:2013mpa}.}; it likewise gives an unbounded path integral in 3+1D. Another parametrization in the literature, $\phi_i(x)=-2\ii\sqrt{c_i+\ii\chi_i(x)}$ \cite{Ogilvie:2008tu}, also gives an unbounded path integral upon closer inspection because there is a flat direction $\chi_i-\chi_{j}$ for $i\neq j$ in the potential. The prior work of Ogilvie and Meisinger on the negative-coupling $\textrm{O}(N)$ model had suggested such a parametrization, missing the fact that it is ill-behaved. Nevertheless, those authors called for ``a detailed understanding of the contours used in
functional integration'' \cite{Ogilvie:2008tu}  and so I have taken steps to be more careful here. As a new result, I propose that the parametrization \eqref{eq:domain-PT} is actually more appropriate. It seems that the kinks are unavoidable.

There is an alternative choice for the domain $\calC_-$ that gets rid of most of these kinks by only forcing one component of the Fourier zero mode of the scalar field to live on the half-lines at $-\pi/4$, $-3\pi/4$ in the lower half of the complex plane. Let $\vec{\phi}(x)=\vec{\phi}_0+\vec{\phi}'(x)$, where $\vec{\phi}_0$ is the zero mode of the scalar field and $\vec{\phi}'$ contains all the non-zero modes. Then the domain $\calC_-$ can be parametrized in terms of the real-valued fields $\chi(x)=\chi_0+\chi'(x)\in\mathbb{R}$ and $\vec{\eta}(x)\in\mathbb{R}^{\Nmin}$ as
\begin{equation}
    \begin{split}
        \vec{\phi}_0\cdot\hat{e}&=
        \chi_0 f(\chi_0), \\
\vec{\phi}'(x)\cdot&\hat{e}=\chi'(x) f(\chi_0),\\
      \vec{\phi}(x)-&\big(\vec{\phi}(x)\cdot\hat{e}\big)\hat{e}=
        \vec{\eta}(x)f(\chi_0),
    \end{split}
    \label{eq:domain-neg-coup}
\end{equation}
where $\chi_0$ and $\chi'$ denote the zero mode and non-zero modes of the field $\chi$, respectively, and where $f$ is the same function defined in \eqref{eq:f}. See figure \ref{fig:PT-domains} for a visualization of this domain and the domain $\calC_{\PT}$ in $\eqref{eq:domain-PT}$. This domain still gives a bounded path integral and respects the $\PT$ symmetry  $\vec{\phi}\to-\vec{\phi}$, $\ii\to-\ii$, but now it makes the path integral amenable to large-$N$ techniques, as we will see. Moreover, it gives a physical partition function.
% \footnote{
However, this domain does not (as far as I can tell) correspond to a Hamiltonian picture where quantum states and an inner product like the $\calC\PT$ inner product can be defined. The domain $\calC_{\PT}$ in equation \eqref{eq:domain-PT} however will have a Hamiltonian, inner product, states, and an associated notion of unitary.
% However, the partition function can be associated with an energy spectrum (see appendix \ref{sec:app:equiv-of-theories}).
% }.
% \footnote{Perhaps this is because the path integral on this domain is ``protected'' in some sense by the $\PT$ symmetry.}. 
For ease of calculation, the domain $\calC_-$ specified by \eqref{eq:domain-neg-coup} is the choice that will be considered in this paper for the negative-coupling theory.

One more remark should be made before continuing on to the calculation of the partition function $Z_-$. It is perhaps unsatisfactory that the domain $\calC_-$ in \eqref{eq:domain-neg-coup}, and also the domain $\calC_{\PT}$ in \eqref{eq:domain-PT}, breaks the usual $\textrm{O}(N)$ symmetry of the theory (which is only a symmetry of the action) down to $\textrm{O}(N-1)$. However, it is not apparent how to keep the $\textrm{O}(N)$ symmetry and the $\PT$ symmetry while keeping the path integral bounded. As was already mentioned, letting the radial coordinate live on the $\PT$-symmetric hyperbola $\vec{\phi}^{\,2}(x)=-4\big(1+\ii \chi(x)\big)$, as was done  for the 1D $N$-component theory in \cite{Bender:2013mpa}, gives an unbounded path integral, even though it keeps the $\textrm{O}(N)$ symmetry. 

It should be pointed out that Ogilvie and Meisinger's original calculation \cite{Ogilvie:2008tu}   using the constraint method (also applied in this paper in subsection \ref{subsec:neg-coup-calc}) was done without \textit{explicitly} breaking the $\textrm{O}(N)$ symmetry. These authors acquired a result equivalent to \eqref{eq:pressure-per-component-simple} in this work, which for the first time I show here to give equivalent thermodynamics to the positive-coupling theory at large $N$. However in using the constraint calculation they do not specify a domain of integration for the field $\vec{\phi}$ or for the auxiliary fields, nor do they explicitly renormalize the theory, write down thermodynamics, or do the Lefschetz thimble analysis that is done later in this work. The only domain they specify, earlier in the work, gives an unbounded path integral and also breaks the $\textrm{O}(N)$ symmetry, and is not used explicitly in the calculation, unlike what will be done with the domain $\calC_-$ here. 

% Alternatively, there are composite $\PT$-symmetric domains that keep the $\textrm{O}(N)$ symmetry, for example by extending the half-lines in the lower half of the complex plane in figure \ref{fig:PT-domains} to full lines, with the sacrifice that the domain of integration intersects itself and is no longer parametrized simply by $N$ real fields. I note that if this is done for the domain $\calC_-$, the $\textrm{O}(N)$ symmetry can be maintained, and one still gets the same result as in this work. One can also restore the $\textrm{O}(N)$ symmetry, allowing it only to be \textit{spontaneously} broken, by integrating the unit vector $\hat{e}$ in \eqref{eq:domain-PT} and \eqref{eq:domain-neg-coup} over the sphere $S^{N-1}$, and one will again get the same result as this work.

Here, I will stick to the domain $\calC_-$ in \eqref{eq:domain-neg-coup} with the intent of demonstrating there is at least \textit{some} domain $\calC_-$ for which the negative-coupling theory appears to ``renormalize into'' the positive-coupling theory.

\subsection{Calculating the partition function \texorpdfstring{$Z_-$}{TEXT}}
\label{subsec:neg-coup-calc}

The partition function $Z_-$ will now be calculated. With the choice in \eqref{eq:domain-neg-coup} for the domain $\calC_-$, and in terms of the real fields $\chi$, $\vec{\eta}$, the path integral \eqref{eq:part-func-neg-coup} for the negative-coupling theory becomes
\begin{equation}
\begin{split}
    Z_- \propto \sum_{+,-}\pm e^{\mp\ii N\pi V\sumintc/4}\bigg(\int_0^{\pm\infty}\upd\chi_0
\int\calD\chi'\,\calD^{\Nmin}\vec{\eta} \,\,e^{-S_-^{\pm}[\chi,\vec{\eta}]}\bigg).
\end{split}
\label{eq:part-func-chi0}
\end{equation}
Here, the symbol $\sumintc$ is shorthand for $\sumintc\equiv\sum_{n}\int \upd^3\mathbf{k}/(2\pi)^3$ (a sum over Matsubara frequencies $\omega_n$ and a integral over 3-momenta $\mathbf{k}$), and $S_-^{\pm}[\chi,\vec{\eta}]$ is given by
\begin{equation}
\begin{split}
        S_-^{\pm}[\chi,\vec{\eta}]=\int_{\beta,V}
    \upd^4x \,\bigg(
    \pm \frac{1}{2\ii}\big((\partial_\mu \chi)^2+(\partial_\mu \vec{\eta})^2 \big)
    +\frac{g}{N} (\chi^2+\vec{\eta}^{\,2})^2\bigg).
\end{split}
\end{equation}
The fields $\chi$, $\vec{\eta}$ have periodic boundary conditions in Euclidean time. We see the path integral becomes a sum of two pieces, each corresponding to one of the half-lines at angles $-\pi/4$, $-3\pi/4$ on which the zero mode $\vec{\phi}_0\cdot\hat{e}$ lives.

In order to solve this integral at large $N$, it is convenient to perform a Hubbard--Stratonovich transformation by introducing the auxiliary fields $\zeta$ and $\sigma=\chi^2+\vec{\eta}^{\,2}$, as follows:
\begin{equation}
\begin{split}
    Z_- \propto \sum_{+,-}\pm e^{\mp\ii N\pi V\sumintc/4}\bigg(\int_0^{\pm\infty}\upd\chi_0
\int\calD\zeta\,\calD\sigma\,\calD\chi'\,\calD^{\Nmin}\vec{\eta} \,\,e^{-S_-^{\pm}[\chi,\vec{\eta},\zeta,\sigma]}\bigg),
\end{split}
\end{equation}
\\
where $S_-^{\pm}[\chi,\vec{\eta},\zeta,\sigma]$ is given by
\begin{equation}
\begin{split}
        S_-^{\pm}[\chi,\vec{\eta},\zeta,\sigma]=\int_{\beta,V}
    \upd^4x \,\bigg(
    \pm \frac{1}{2\ii}\big((\partial_\mu \chi)^2+(\partial_\mu \vec{\eta})^2 \big)
    +\frac{g}{N} \sigma^2+\frac{\ii}{2}\zeta(\chi^2+\vec{\eta}^{\,2}-\sigma)\bigg).
\end{split}
\label{eq:action-PT-Hubbard-Stratonovich}
\end{equation}
This amounts to the insertion of a delta function into the path integral that enforces $\sigma=\chi^2+\vec{\eta}^{\,2}$. The action \eqref{eq:action-PT-Hubbard-Stratonovich} is quadratic in the auxiliary field $\sigma$, so $\sigma$ can be integrated out to give
\begin{equation}
\begin{split}
    Z_- \propto \sum_{+,-}\pm e^{\mp\ii N\pi V\sumintc/4}\bigg(\int_0^{\pm\infty}\upd\chi_0
\int\calD\zeta\,\calD\chi'\,\calD^{\Nmin}\vec{\eta} \,\,e^{-S_-^{\pm}[\chi,\vec{\eta},\zeta]}\bigg),
\end{split}
\end{equation}
where $S_-^{\pm}[\chi,\vec{\eta},\zeta]$ is now given by
\begin{equation}
\begin{split}
        S_-^{\pm}[\chi,\vec{\eta},\zeta]=\int_{\beta,V}
    \upd^4x \,\bigg(
    \pm \frac{1}{2\ii}\big((\partial_\mu \chi)^2+(\partial_\mu \vec{\eta})^2 \big)
    +\frac{\ii}{2}\zeta(\chi^2+\vec{\eta}^{\,2} )+\frac{N\zeta^2}{16g}\bigg).
\end{split}
\label{eq:action-post-Hubbard-Stratonovich}
\end{equation}

The auxiliary field $\zeta$ can be split into a zero mode $\zeta_0$ and non-zero modes $\zeta'$ as $\zeta(x)=\zeta_0+\zeta'(x)$. At leading order in large $N$, only the zero mode contributes to the pressure\footnote{Including the non-zero modes $\zeta'$ amounts to including $1/N$ corrections, and can be done via R$n$ resummation methods \cite{Romatschke:2019rjk,Romatschke:2019wxc}.}, so the integral over the non-zero modes can be neglected \cite{Romatschke:2019ybu} and the partition function becomes
\begin{equation}
\begin{split}
    Z_- \propto \sum_{+,-}\pm e^{\mp\ii N\pi V\sumintc/4}\bigg(\int_0^{\pm\infty}\upd\chi_0\int_{-\infty}^{\infty}\upd\zeta_0\int\calD\chi'\calD^{\Nmin}\vec{\eta} \,\,e^{-S_-^{\pm}[\chi, \vec{\eta},\zeta_0]}\bigg),
\end{split}
\label{eq:part-func-auxiliary-zero-mode}
\end{equation}
with $\zeta$ in expression \eqref{eq:action-post-Hubbard-Stratonovich} now replaced by only the zero mode $\zeta_0$. 

There is a  term $\ii\zeta_0(\chi^2+\eta^2)/2$ in the action $S_-^{\pm}[\chi, \vec{\eta},\zeta_0]$ in \eqref{eq:part-func-auxiliary-zero-mode} which contains a cross-term $\ii\zeta_0\chi_0\chi'$ between the zero and non-zero modes of $\chi$. Because $\zeta_0\chi_0$ does not vary with $x$ and $\chi'$ contains only non-zero modes, this cross-term vanishes under the integral $\int_{\beta,V}\upd x^4$, leaving one with a term $\ii\zeta_0(\chi_0^2+\chi'^2+\eta^2)/2$ in the action. The action $S_-^{\pm}[\chi, \vec{\eta},\zeta_0]$ is thus an even function of $\chi_0$, and so the bounds of the integral over $\chi_0$ can be extended from $[0,\pm\infty)$ to $(-\infty,\infty)$ without consequence, which gives
\begin{equation}
\begin{split}
    Z_- \propto \sum_{+,-} e^{\mp\ii N\pi V\sumintc/4}\int_{-\infty}^{\infty}\upd\zeta_0\int\calD^N\vec{\chi}\,\,e^{-S_-^{\pm}[\vec{\chi},\zeta_0]},
\end{split}
\end{equation}
where $\vec{\chi}(x)=\big(\chi(x),\vec{\eta}(x)\big)\in\mathbb{R}^N$ has been introduced simply to group together the fields $\chi(x)$ and $\vec{\eta}(x)$, and $S_-^{\pm}[\vec{\chi},\zeta_0]$ is
\begin{equation}
    \begin{split}
        S_-^{\pm}[\vec{\chi},\zeta_0]=\beta V\frac{N\zeta_0^2}{16g}
        +\int_{\beta,V} \upd x^4 \bigg(\pm\frac{1}{2\ii}(\partial_\mu\vec{\chi})^2+\frac{\ii}{2}\zeta_0\vec{\chi}^{\,2}\bigg).
    \end{split}
    \label{eq:action-auxiliary-zero}
\end{equation}

The action \eqref{eq:action-auxiliary-zero} is quadratic in $\vec{\chi}$, and so $\vec{\chi}$ can be integrated out, yielding
\begin{equation}
\begin{split}
        Z_- \propto \,\,  \sum_{+,-} e^{\mp\ii N\pi V\sumintc/4}\bigg(&\int_{-\infty}^{\infty}\upd\zeta_0
        \,e^{-N\ln\det(\pm\ii\partial_\mu^2+\ii\zeta_0)/2-\beta V N\zeta_0^2/16g}\bigg)\\
        \propto \,\, &\sum_{+,-} \int_{-\infty}^{\infty}\upd\zeta_0\, e^{-N\tr\ln(-\partial_\mu^2\mp\zeta_0)/2-\beta V N\zeta_0^2/16g}\\
        =\,\,&\sum_{+,-} \int_{-\infty}^{\infty}\upd\zeta_0\,e^{N\beta V p(\beta,\mp\zeta_0)}.
\end{split}
\label{eq:integral-zeta0}
\end{equation}
Note that the complex Jacobian term $e^{\mp\ii N\pi V\sumintc/4}$ has been canceled by pulling out a $e^{-N\tr\ln(\mp \ii)/2}$ from inside the path integral, where the trace of the operator on a single field component brings out a phase space factor $V \sumintc$. Here, also, I've introduced the pressures per component $p(\beta,\mp\zeta_0)$, which are given more explicitly by
\begin{equation}
\begin{split}
        p(\beta,\mp\zeta_0)=-\frac{\zeta_0^2}{16g}-
    \frac{1}{2\beta}\sum_{n=-\infty}^{\infty}\int \mu^{2\epsilon}\frac{\upd^{3-2\epsilon} \mathbf{k}}{(2\pi)^3}\ln(\omega_n^2+\mathbf{k}^2\mp\zeta_0),
\end{split}
\label{eq:pressure-per-component-simple}
\end{equation}
where $\omega_n=2\pi n/\beta$ is the $n$th Matsubara frequency and the integral over 3-momenta $\mathbf{k}$ can be done using dimensional regularization in $3-2\epsilon$ dimensions. $\mu$ is the scale of dimensional regularization, related to the $\overline{\textrm{MS}}$ scale $\mubar$.  

The thermal sum-integral in \eqref{eq:pressure-per-component-simple} has been done, for example, in \cite{Laine:2016hma}, and the resulting pressures per component are
\begin{equation}
\begin{split}
    p(\beta,\mp\zeta_0)= -\frac{\zeta_0^2}{16g}+\frac{\zeta_0^2}{64\pi^2}\bigg(\frac{1}{\epsilon}+\ln\bigg(\frac{\mubar^2}{\mp \zeta_0}\bigg)+\frac{3}{2}\bigg)
    \mp\frac{\zeta_0}{2\pi^2\beta}\sum_{n=1}^\infty \frac{K_2(n\beta\sqrt{\mp\zeta_0})}{n^2}.
\end{split}
\end{equation}
This is the same expression as \eqref{eq:p-per-component-pos-coup} from the positive-coupling theory, with $m^2$ replaced by $m^2=\mp\zeta_0$ and $\lambda$ replaced by $-g$.

As with the positive-coupling theory, there is a $1/\epsilon$ divergence from the dimensional regularization that can be absorbed by defining the renormalized coupling $\gR(\mubar)$ as
\begin{equation}
    \frac{1}{\gR(\mubar)}=\frac{1}{g(\mubar)}-\frac{1}{4\pi^2\epsilon}.
\end{equation}
Then, requiring the pressures to be independent of the $\overline{\textrm{MS}}$ scale $\mubar$, such that $\partial p(\beta,\mp\zeta_0)/\partial\mubar=0$, yields the beta function and fixes the running coupling to be
\begin{equation}
    \gR(\mubar)=\frac{4\pi^2}{\ln(\mubar^2/\LambdaLP^2)},
    \label{eq:running-neg-coupling}
\end{equation}
where $\LambdaLP$ is again a scale introduced from the integration of the beta function.
The renormalized pressures $p(\beta,\mp\zeta_0)$ are then given by
\begin{equation}
    \begin{split}
    p(\beta,\mp\zeta_0)=\frac{\zeta_0^2}{64\pi^2}\bigg(\ln\bigg(\frac{\LambdaLP^2}{\mp\zeta_0}\bigg)+\frac{3}{2}\bigg)
           \mp\frac{\zeta_0}{2\pi^2\beta^2}\sum_{n=1}^{\infty} \frac{K_2(n\beta \sqrt{\mp\zeta_0})}{n^2}.
\end{split}
\label{eq:pressure-renormed-neg-coup}
\end{equation}

\begin{figure}
    \centering
    \includegraphics[width=68mm]{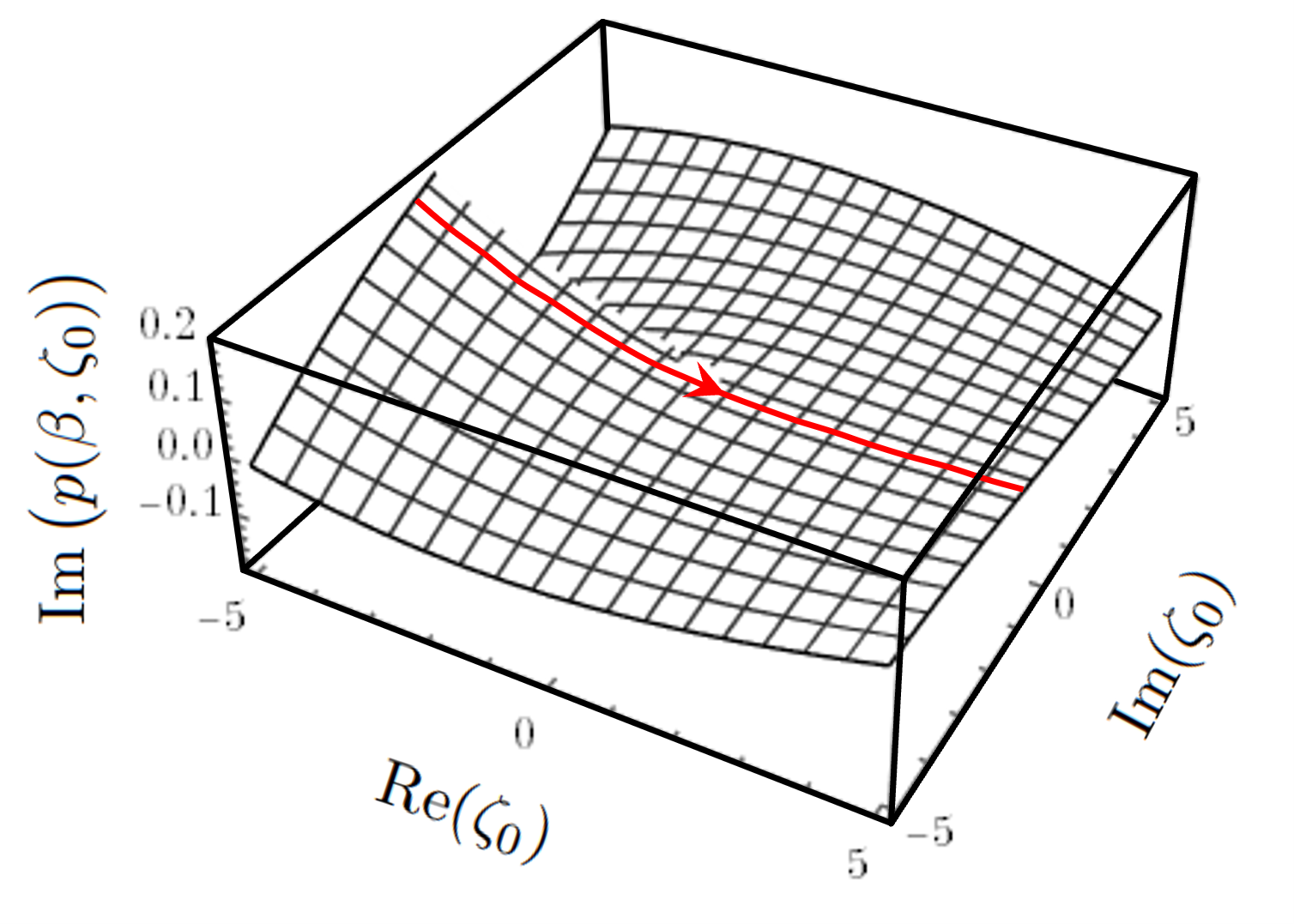}
    \caption{The imaginary part of the pressure $p(\beta,\zeta_0)$ in \eqref{eq:pressure-renormed-neg-coup} at a temperature $T=1/\beta=0.65\,\LambdaLP$, showing that the function is multi-sheeted. The horizontal axes are expressed in units of $\LambdaLP^2$ and the vertical axis in units of $\LambdaLP^4$. There is a branch point at $\zeta_0=0$ and a branch cut can be made along the negative half of the real axis. In order to avoid the branch cut, one has to integrate $e^{N\beta V p(\beta,\zeta_0)}$ along a contour $\zeta_0+\ii0^+\in\mathbb{R}$ slightly shifted in the imaginary direction, which is shown in red.}
    \label{fig:Riemann-sheet}
\end{figure}

Note that the running coupling in \eqref{eq:running-neg-coupling} is exactly the same expression as the running coupling in \eqref{eq:running-coupling} for the positive-coupling theory up to a minus sign. That is, $\gR(\mubar)=-\lambdaR(\mubar)$ if the scale $\LambdaLP$ in both theories is identified as the same scale. 

At large $N$, the sum of integrals in \eqref{eq:integral-zeta0} can be replaced by
\begin{equation}
    Z_-\approx\sum_{+,-}n_j^\pm\sum_{j} e^{N\beta V p(\beta,\zeta_{0j})},
    \label{eq:thimbles-saddle-point}
\end{equation}
where $\zeta_{0j}$ denotes the $j$th saddle of $p(\beta,\zeta_0)$ and $n^\pm _j$ denotes the intersection number of the contour of integration with the Lefschetz anti-thimble passing through the $j$th saddle. For an introduction to Lefschetz thimbles, see \cite{Witten:2010cx}. Only the relevant saddles with $n_j\neq 0$ contribute to the partition function, and only the saddle with the largest pressure dominates in the large-volume limit. 
In order to determine the saddles,
one has to apply the saddle condition (or gap equation) $\partial p(\beta,\mp\zeta_0)/\partial\zeta_0=0$ to fix the gap $m=\sqrt{\mp\zeta_0}$. 

However, the function for the pressures in \eqref{eq:pressure-renormed-neg-coup} has a branch point at $\zeta_0=0$ and is multi-sheeted. See figure \ref{fig:Riemann-sheet} for an illustration of this feature. A branch cut can be made on the real axis for $\mp\zeta_0<0$, defining the principle sheet, and the integrals of $e^{N\beta V p(\beta,\mp\zeta_0)}$ in \eqref{eq:integral-zeta0} can be done slightly above or below the real axis in order to avoid the branch point. Here there is an apparent ambiguity whether the integration should occur below or above the branch point, but the requirement of $\PT$ symmetry settles the ambiguity. Since $\PT$ symmetry interchanges $\zeta_0\to-\zeta_0$ and $\ii\to-\ii$ \footnote{This is because $\PT$ exchanges the two half lines at angles $-\pi/4$, $-3\pi/4$ in figure \ref{fig:PT-domains}, or equivalently switches the two integrals over $\chi_0>0$ and $\chi_0<0$ in \eqref{eq:part-func-chi0}, which in \eqref{eq:integral-zeta0} amounts to swapping $p(\beta,-\zeta_0)$ with $p(\beta,\zeta_0)$, i.e. $\zeta_0\to-\zeta_0$.}, the integrals in \eqref{eq:integral-zeta0} should be
\begin{equation}
\begin{split}
       Z_-\propto&\int_{-\infty}^{\infty} \upd\zeta_0\,e^{N\beta V p(\beta,-\zeta_0+\ii0^+)}
       +\int_{-\infty}^{\infty} \upd\zeta_0\,e^{N\beta V p(\beta,\zeta_0-\ii0^+)}\\
       =&\int_{-\infty}^{\infty} \upd\zeta_0\,e^{N\beta V p(\beta,\zeta_0+\ii0^+)}
       +\int_{-\infty}^{\infty} \upd\zeta_0\,e^{N\beta V p(\beta,\zeta_0-\ii0^+)}
\end{split}
\label{eq:part-func-zeta0}
\end{equation}
where here I have used the fact that $\int_{-a}^{a} \upd x \,f(-x)=\int_{-a}^{a} \upd x \,f(x)$ for any function $f$. 

\begin{figure}
    \centering
    \includegraphics[width=70mm]{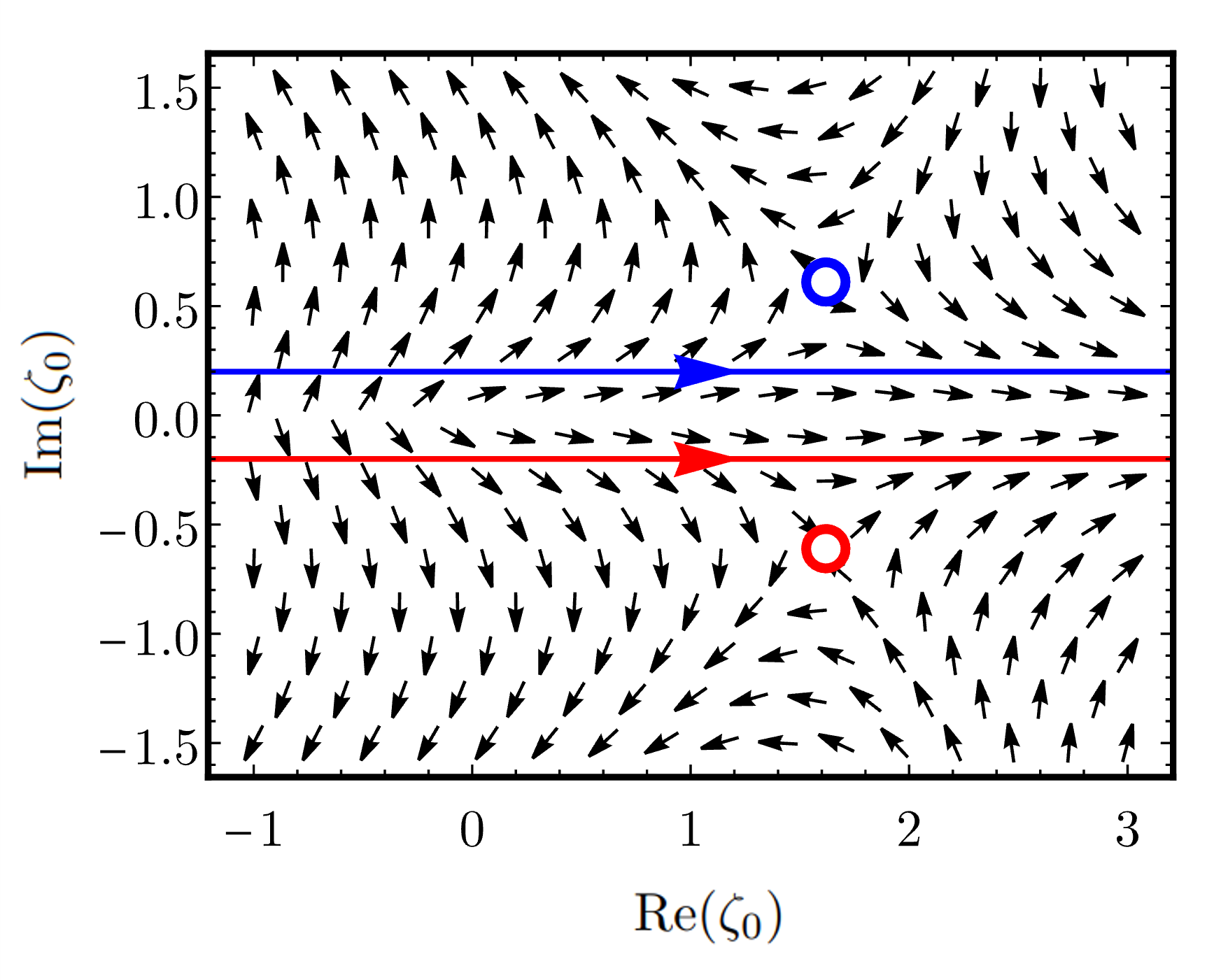}
    \caption{The downward flow $-\big(\partial p(\beta,\zeta_0)/\partial\zeta_0\big)^*$ of the pressure $p(\beta,\zeta_0)$ at a temperature $T=1/\beta=0.65\,\LambdaLP$. Both axes are expressed in units of $\LambdaLP^2$. The line of integration of $e^{N\beta V p(\beta,\zeta_0)}$ over $\zeta_0+\ii0^+\in\mathbb{R}$ is indicated in red. Note that it picks up a contribution only from the lower saddle indicated in red. The line of integration $\zeta_0-\ii0^+\in\mathbb{R}$ in blue corresponds to the integral of $e^{N\beta V p(\beta,-\zeta_0)}$, and it picks up a contribution only from the upper saddle indicated in blue. The total partition function is a sum of both integrals, so it gets a contribution from both complex conjugate saddles for $T>\Tc$.}
    \label{fig:downward-flow}
\end{figure}

The relevant saddles of $p(\beta,\zeta_0)$ in \eqref{eq:part-func-zeta0} can be determined from the downward flow of the contours $\zeta_0\pm\ii0^+\in\mathbb{R}$ onto the Lefschetz thimbles (contours of steepest descent) passing through the saddles. The downward flow is defined as
\begin{equation}
    \pdrv{\zeta_0}{s} = - \bigg(\pdrv{p(\beta,\zeta_0)}{\zeta_0}\bigg)^*,
\end{equation}
where $s$ is ``flow time'' and the the lines following the downward flow toward the saddle are the Lefschetz anti-thimbles. A visualization of this downward flow at one selected temperature is in figure \ref{fig:downward-flow}. 

For $T<\Tc\approx0.616\,\LambdaLP$ there are two real saddles which are relevant for both contours $\zeta_0\pm\ii0^+\in\mathbb{R}$, and the saddle with the larger value of $\zeta_0>0$ gives the dominant pressure per component $p(\beta)$ in \eqref{eq:large-N-Z}. At zero temperature this saddle has $\zeta_0=e\LambdaLP^2$, which corresponds to a non-vanishing mass gap $m=\sqrt{\zeta_0}$ for the scalar field $\vec{\phi}$ in the vacuum. The saddle of the non-preferred phase is trivial at $T=0$. When the scale $\LambdaLP$ is equated to the scale $\LambdaLP$ in the positive-coupling theory, these saddles correspond exactly to those for the positive-coupling theory and those found by Romatschke in \cite{Romatschke:2022jqg}, and the pressures of these saddles are the same in both theories.  Above the critical temperature $T>\Tc$, there are, just as in the positive-coupling theory, a complex conjugate pair of saddles. The second integral in \eqref{eq:part-func-zeta0}, over the contour slightly below the branch cut (shown in figures \ref{fig:Riemann-sheet} and \ref{fig:downward-flow}), picks up a contribution only from the saddle with $\Im(\zeta_0)<0$. Meanwhile, the first integral, on the contour above the branch cut in figure \ref{fig:downward-flow}, picks up a contribution from the saddle with $\Im(\zeta_0)>0$. The saddles have the same pressures $p_+(\beta)$, $p_-(\beta)=p^{*}_+(\beta)$ as the complex conjugate pair of saddles in the positive-coupling theory. Just as in \eqref{eq:large-N-Z-2}, keeping only the part of $\ln Z_-$ that scales like $N$ gives a pressure per component $p(\beta)$ in \eqref{eq:large-N-Z} that is equal to $\Re p_+(\beta)$. 

This concludes the calculation of the thermal partition function $Z_-$ for the $-g\vec{\phi}^4/N$ theory at large $N$. The pressure per component $p(\beta)$ as a function of temperature is plotted in figure \ref{fig:pressure}.

One last thing to recall is that the domain $\calC_{-}$ in \eqref{eq:domain-neg-coup} broke the $\textrm{O}(N)$ symmetry. Alternatively to $\calC_{\PT}$ or $\calC_{-}$, there are composite $\PT$-symmetric domains that would have kept the $\textrm{O}(N)$ symmetry, for example by extending the half-lines in the lower half of the complex plane in figure \ref{fig:PT-domains} to full lines, with the sacrifice that the domain of integration intersects itself and is no longer parametrized simply by $N$ real fields. I note that if this is done for the domain $\calC_-$, the $\textrm{O}(N)$ symmetry can be maintained, and it is fairly easy to see that one still gets the same results as in this section. One can also restore the $\textrm{O}(N)$ symmetry, allowing it only to be \textit{spontaneously} broken, by integrating the unit vector $\hat{e}$ in \eqref{eq:domain-PT} and \eqref{eq:domain-neg-coup} over the sphere $S^{N-1}$, and one will again get the same result as this work. Therefore the breaking of the $\textrm{O}(N)$ symmetry down to $\textrm{O}(N-1)$ by the domain $\calC_{-}$ of path integration appears not to have an observable impact.

I note that the path integral \eqref{eq:part-func-neg-coup} yields different results if evaluated over different Stokes sectors, which are the regions in which the path integral domain can terminate to give a bounded result. For example, if the path-integration domain is specified by $N$ real fields $\vec{\chi}(x)\in\mathbb{R}^N$ as $\vec{\phi}=\vec(\chi) e^{\ii\pi/4}$, this corresponds to selecting a non-$\PT$-symmetric set of Stokes sectors. In this case, carrying through the calculation in this section, including the Lefschetz thimble analysis, will yield a partition function that above $T=\Tc$ has a single saddle with a complex pressure with $\Im p(\beta)\neq 0$, $\beta<\beta_C$. The partition function will then be complex at high temperatures. This follows from the Lefschetz thimble analysis, in which $\PT$ symmetry of the domain $\calC_{-}$ was important in selecting both complex-conjugate saddles (rather than just one as in this case). Thus the choice of Stokes sectors (namely, that they be $\PT$-symmetric) is relevant.

I discuss results in the next section.

%% file: results.tex
\section{Results}
\label{sec:results}

The primary thing to observe from the preceding calculation of $Z_-$ is the following: At large $N$, when the scale $\LambdaLP$ appearing in both the positive- and negative-coupling $\vec{\phi}^4$ theories is considered to be the same scale, the thermal partition functions of both theories are equal, $Z_+=Z_-$, and the renormalized couplings $\lambdaR(\mubar)$ and $\gR(\mubar)$ are simply related by a minus sign. In other words, there exists a complex domain of path integration $\calC_-$ for the $-g\vec{\phi}^4/N$ theory such that
\begin{equation}
    Z_-=\int_{\calC_-} \calD^N\vec{\phi}\, e^{-S_-[\vec{\phi}]}=\int \calD^N\vec{\phi}\, e^{-S_+[\vec{\phi}]}=Z_+
\end{equation}
when the renormalized couplings in both theories are identified as $\lambdaR(\mubar)=-\gR(\mubar)$. The two theories are only different in the range of choices for the $\overline{\textrm{MS}}$ renormalization scale $\mubar$. For the $\lambda\vec{\phi}^4/N$ theory, $\mubar$ is taken to be less than $\LambdaLP$, whereas for the $-g\vec{\phi}^4/N$ theory, $\mubar$ is taken to be greater than $\LambdaLP$. However, the result is independent of the choice of $\mubar$, and therefore independent of the sign of the coupling. Thus we have seen a calculation for $\mubar>\LambdaLP$ in which a thermodynamic observable (the pressure $p(\beta)$) after renormalization is no different than when calculated with $\mubar<\LambdaLP$. The idea that observables are cutoff-independent is the philosophy of renormalization. This, I believe, strongly suggests that the negative-coupling theory is something like the UV completion of the positive-coupling theory, and that perhaps they are one and the same theory. In that case, the $\vec{\phi}^4$ theory at large $N$ has a negative bare coupling constant, as argued early on in \cite{Kobayashi:1975ev} and later in \cite{Romatschke:2022jqg,Romatschke:2022llf,Romatschke:2023sce}, and it would seem that the sign of the coupling changes under renormalization.

Note that the $\PT$ symmetry of the domain $\calC_-$ was important in ensuring this equivalence between partition functions, at least above the temperature $\Tc$. (Namely, the $\PT$ symmetry played a role in the Lefschetz thimble analysis for the integral \eqref{eq:integral-zeta0} over the auxiliary field $\zeta_0$.)

% This involves identifying the scale $\LambdaLP$ of the Landau pole in both theories as one and the same scale.

\begin{figure}
    \centering
    \includegraphics[width=75mm]{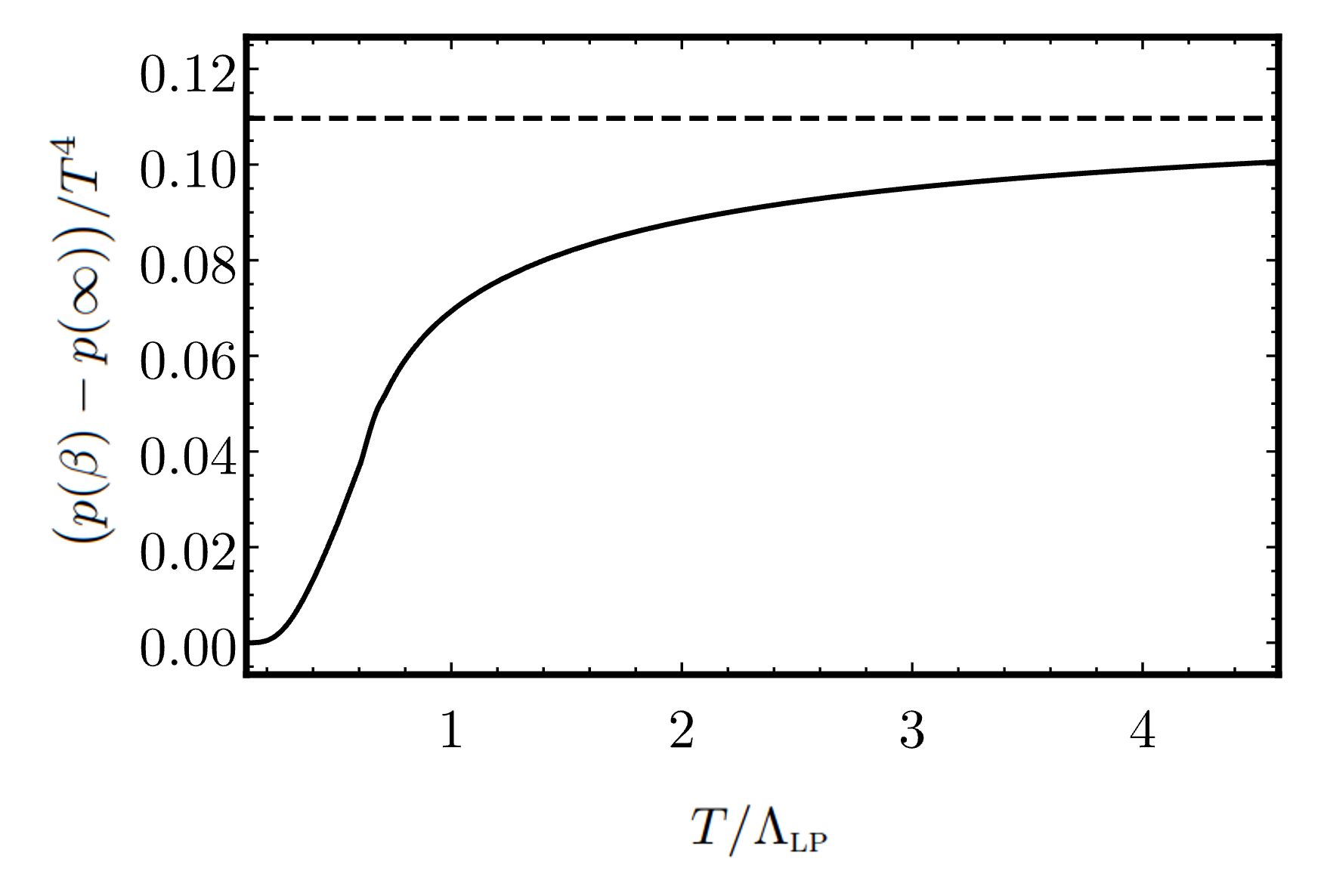}
    \caption{A plot of the pressure per component $p(\beta)$ as a function of temperature $T=1/\beta$ for the $-g\vec{\phi}^4/N$ theory. The vacuum pressure is subtracted and $p(\beta)$ is scaled by $T^4$. This agrees with the result of Romatschke in \cite{Romatschke:2022jqg}. The dashed line indicates the Stefan--Boltzmann limit $p(\beta)=\pi^2 T^4/90$ for a free scalar boson. One can see that the pressure approaches the free theory limit at high temperatures, which is a consequence of asymptotic freedom.}
    \label{fig:pressure}
\end{figure}

 The pressure per component $p(\beta)$ of the $\vec{\phi}^4$ theory with negative bare coupling is plotted in figure \ref{fig:pressure} and is identical to that found by Romatschke in \cite{Romatschke:2022jqg}. At $T=\Tc$ the volumetric heat capacity $c_V(T)=N T \partial^2 p(\beta)/\partial T^2$ is discontinuous and so there is a 2nd-order phase transition.

\subsection{Non-triviality and asymptotic freedom}

A notable feature of the pressure in figure \ref{fig:pressure} is that it approaches the Stefan--Boltzmann limit $p(\beta)=\pi^2 T^4/90$ for a free scalar boson at high temperatures. This is a characteristic feature of asymptotically free theories. This makes perfect sense from having a negative bare coupling $\lambda_0=-g_0<0$ as $\Lambda_0\to\infty$ as seen in figure \ref{fig:Landau-pole}; the sign of the beta function $\partial \gR/\partial \ln(\mubar)$ is opposite the sign of the coupling as $\mubar\to\infty$ and the coupling \eqref{eq:running-neg-coupling} approaches zero in the UV, so that the theory is almost free at high energies. 

In doing the calculation of the partition function $Z_-$ with negative coupling, the only trouble we ran into was that the scale $\mubar$ could not be taken below the scale $\LambdaLP$ or else the assumption of negative coupling would be violated\footnote{If the coupling were taken to be positive, as happens for $\mubar<\LambdaLP$, the path integral \eqref{eq:part-func-neg-coup} on the domain $\calC_-$ would be unbounded.}. Thus the negative-coupling theory has a Landau pole, albeit a Landau pole\footnote{Here, by Landau pole, I mean any place where the coupling diverges and becomes infinite.} in the IR and not in the UV. 
% Identifying this Landau pole with the Landau pole in the UV of the positive-coupling theory shows that the two theories
A Landau pole in the IR, unlike in the positive-coupling case, does not prevent there
from being an interacting continuum limit of the theory. Indeed, the negative-coupling $\vec{\phi}^4$ theory at large $N$ is \textit{not} quantum trivial. 

But we have seen here that the negative-coupling theory appears to ``extend'' the positive-coupling theory to scales $\mubar$ greater than that of the Landau pole $\LambdaLP$, giving an equivalent partition function $Z_+=Z_-$ at any scale $\mubar$. As mentioned before, renormalization describes an equivalence of observables under changes of the cutoff scale and bare coupling. Here we've taken the coupling negative, and the scale $\mubar$ greater than $\LambdaLP$, and found no change in observable thermodynamics. This may mean that an interacting continuum limit of the positive-coupling theory \textit{can} be taken while keeping $\LambdaLP$ finite. The continuum limit $\mubar>\LambdaLP$ in that case would simply be the negative-coupling theory. Therefore, since the negative-coupling theory is non-trivial, that would mean the standard $\vec{\phi}^4$ theory at large $N$ is non-trivial, as pointed out by Romatschke \cite{Romatschke:2023sce}. I leave it to future work to establish if this is truly the case.

\subsection{High temperatures \texorpdfstring{$T>\Tc$}{TEXT}}
\label{subsec:high-temps}

One last result should be mentioned. The appearance of two saddles with complex-conjugate pressures $p_+(\beta),p_+^*(\beta)$ for temperatures $T>\Tc$ above the critical temperature $\Tc$ is reminiscent of a phase of spontaneously broken $\PT$ symmetry, which can occur in non-Hermitian Hamiltonian systems when energy eigenvalues appear in complex-conjugate pairs \cite{Bender:1998gh}, causing the partition function to become oscillatory as a function of temperature. I (briefly) provide further explanation to those that may be unfamiliar with the $\PT$ symmetry literature: In a $\PT$-symmetric non-Hermitian Hamiltonian system, spontaneous $\PT$ symmetry breaking gives a thermal partition function $Z(\beta)$ of the form
\begin{equation}
    Z(\beta)=\sum_{n=1}^{N_{\textrm{R}}} e^{-\beta E_{\textsc{r},n}} + \sum_{n=1}^{N_{\textrm{C}}} \big(e^{-\beta E_{\textsc{c},n}}+e^{-\beta E^*_{\textsc{c},n}}\big)
    \label{eq:osc-part}
\end{equation}
where the Hamiltonian has eigenvalues $E_{\textsc{r},n}\in\mathbb{R}$ that are real, as well as eigenvalues  $E_{\textsc{c},n},\in\mathbb{C},\notin\mathbb{R}$ which come in complex-conjugate pairs. The partition function then has terms that oscillate with temperature $T=1/\beta$ as $\sim2 e^{-\beta \Re E_{\textsc{c},n}} \cos(\beta \Im E_{\textsc{c},n})$. Using standard large-$N$ methods we see this kind of behavior at high temperatures $T\gtrsim 0.616\, \LambdaLP$ in the $\textrm{O}(N)$ model here.

 In the formal large-$N$ limit a physical pressure per component $p(\beta)$ in \eqref{eq:large-N-Z} can still be extracted, but in this case the appearance of complex saddles signals a \textit{dynamical} instability. This instability might affect real-time dynamics (for example, in calculations of viscosities or other transport coefficients along the lines of \cite{Romatschke:2021imm,Weiner:2022kgx,Lawrence:2022vwa}), since a complex conjugate pair of masses $m^2$ for the field $\vec{\phi}$ leads to an unstable pole of the propagator in the region $\Im(\omega)>0$ when Wick-rotated into real time. However, I note that oscillatory partition functions as in \eqref{eq:osc-part} or appearances of complex saddles have been discussed in previous literature on $\PT$ symmetry (e.g. \cite{Meisinger:2012va}) and on $\mathcal{C}\mathcal{K}$ symmetry (a type of $\PT$ symmetry) in finite-density quantum chromodynamics \cite{Nishimura:2014kla}, and it is not clear that they de facto are ``sick''. But in our case here there is a dynamical instability due to the complex masses $m,\,m^*$, and that poses a problem.

Even considering this dynamical instability, it is too soon to conclude that the theory is unstable or somehow sick at high temperatures (just as it was prematurely concluded in the 70s that the positive-coupling theory was ``sick'' at zero temperature \cite{PhysRevD.10.2491,PhysRevD.13.2212}). This is because there may be non-constant (i.e. ``inhomogeneous'') saddles $\zeta(x)$ of the action for the auxiliary field (once $\vec{\phi}$ is integrated out) that are preferred at high temperatures, so that this apparent $\PT$-broken phase is suppressed in the large-volume limit. 

% Admittedly, the relevance of these non-constant saddles might affect the result of this work. However, at
If this is the case, at leading order in large $N$ there is still a reason to expect that the partition functions $Z_+$ and $Z_-$ of the positive- and negative-coupling theories will be identical after renormalization. This can be shown straightforwardly. The partition function $Z_+$ allowing for a non-constant gap parameter $m(x)=\sqrt{\ii\zeta(x)}$ is
\begin{equation}
\begin{split}
          Z_+ \propto \int \calD\zeta\,e^{-N\tr\ln(-\partial_\mu^2+\ii\zeta)/2-\int_{\beta,V} N\zeta^2/16\lambda}
          = \int \calD\zeta\, e^{-N s[\ii\zeta]},
\end{split}
\end{equation}
where $\int_{\beta,V}$ is shorthand for $\int_{\beta,V} \upd^4 x$ and I've introduced the action per component 
\begin{equation*}
    s[m^2]=\frac{1}{2}\tr\ln\big(-\partial_\mu^2+m^2(x)\big)+\int_{\beta,V}\upd^4 x \frac{m^4}{16\lambda}.
\end{equation*}
At leading order in large $N$ this will be replaced by a sum over the relevant saddles $\zeta(x)=-\ii m^2(x)$ which satisfy the condition $\delta s[m^2]/\delta m^2(x)=0$, i.e.
\begin{equation}
     Z_+\approx n_j\sum_{j} e^{-N s[m^2_j]},
     \label{eq:saddles-pos-coup}
\end{equation}
with $m^2_j$ denoting the $j$th saddle, and $n_j$, as in \eqref{eq:thimbles-saddle-point}, denoting the intersection number of the domain of path integration with the Lefschetz anti-thimble passing through the $j$th saddle. Meanwhile the partition function of the negative-coupling theory is given by
\begin{equation}
\begin{split}
          Z_- \propto &\sum_{+,-}\bigg(\int \calD\zeta
          e^{-N\tr\ln(-\partial_\mu^2\mp\zeta)/2-\int_{\beta,V} N\zeta^2/16g+\Delta\ln Z_-^\pm[\zeta]}\bigg)\\
          &\propto \sum_{+,-}\pm\int\calD\zeta\,e^{-N s[\mp\zeta]+\Delta\ln Z_-^\pm[\zeta]},
\end{split}
\label{eq:integral-zeta-non-const}
\end{equation}
where there is a $\mathcal{O}(N^0)$ term $\Delta\ln Z_-^{\pm}[\zeta]$ which is given by
\begin{equation}
\begin{split}
          e^{\Delta \ln Z^\pm_-[\zeta]} \propto 
          \pm\int_0^{\pm\infty}\upd\chi_0
\int\calD\chi'\,\,e^{-\Delta S_-^{\pm}[\chi,\zeta]}
\end{split}
\end{equation}
with
\begin{equation}
\begin{split}
          \Delta S_-^{\pm}[\chi,\zeta]=-\frac{1}{2}\tr\ln\big(\pm\ii\partial_\mu^2+\ii\zeta(x)\big)
          +\int_{\beta,V} \upd^4 x\, \bigg(\pm \frac{1}{2\ii}(\partial_\mu\chi)^2+\frac{\ii}{2}\zeta\chi^2\bigg).
\end{split}
\end{equation}
At leading order in large $N$, this term  does not contribute 
and the path integral can also be replaced by a sum over saddles of $s[m^2]$:
\begin{equation}
      Z_-\approx \sum_{+,-} n^{\pm}_j\sum_{j} e^{-N s[m^2_j]},
\label{eq:saddles-neg-coup}
\end{equation}
with $n^\pm_j$ and $m^2_j=\mp\zeta_j$ defined similarly as in \eqref{eq:thimbles-saddle-point}. In \eqref{eq:integral-zeta-non-const} and \eqref{eq:saddles-neg-coup}, the functional $s[m^2]$ is the same as for the positive-coupling theory, with $\lambda$ replaced by $-g$. Comparing \eqref{eq:saddles-pos-coup} and \eqref{eq:saddles-neg-coup}, one can see that the partition functions at positive and negative coupling will be equivalent at large $N$ (when the scale $\LambdaLP$ is equivalent) as long as there are non-zero intersection numbers $n_j$ or $n^\pm_j$ for the same dominant saddles $m^2_j$ in both the positive- and negative-coupling theories.

If the non-constant saddles $\zeta(x)$ dominate at high temperatures, however, I might expect that the 2nd-order phase transition would become a 1st-order phase transition. This should be investigated.

\subsection{Possible objections}

One might object to the presence of asymptotic freedom in large-$N$ scalar and fermionic theories and in $\PT$-symmetric theories. Coleman and Gross in the 70s argued that the only asymptotically free theories are non-Abelian gauge theories. However, their argument assumed that scalar field theory with a negative coupling constant led to an unbounded potential \cite{PhysRevLett.31.851}. It is no longer true that the potential is unbounded from below if the scalar fields live on an appropriate complex domain, such as $\calC_{\PT}$ in \eqref{eq:domain-PT} or $\calC_-$ in \eqref{eq:domain-neg-coup}, or the $\PT$-symmetric domain considered in \cite{Bender:2006wt,Jones:2006qs} for the $-g\phi^4$ theory. 
Thus scalar field theories can in fact be asymptotically free.

An objection can also be raised specifically about the $\vec{\phi}^4$ model at large $N$. In the 70s, Coleman, Jackiw, and Politzer studied the $\textrm{O}(N)$ model at large $N$ in 3+1D and found it possesses a tachyon, and on this basis they argued that the theory is ``sick'' \cite{PhysRevD.10.2491}. However, it must be pointed out that this tachyon appears only in the dispreferred vacuum in which the field $\vec{\phi}$ has no mass gap. This instability goes away when one realizes that the correct vacuum is the one where the field has a mass gap $m=\sqrt{e}\LambdaLP$, as pointed out by Abbott, Kang, and Schnitzer \cite{PhysRevD.13.2212}.

%% file: conclusion.tex
\section{Conclusion}

The $\vec{\phi}^4$ theory is just one recent example of large-$N$ theories with negative or complex bare couplings. Berges, Gurau, and Preis studied an $\textrm{O}(N)^3$ model with one coupling $\ii g$ that is imaginary in the UV and possesses a Landau pole in the IR,  which also exhibits asymptotic freedom \cite{Berges:2023rqa}. Grable and Weiner studied a Gross--Neveu-like model of $N$ interacting fermions in 3+1D, and found that the coupling diverges at a scale $\Lambda_{\textsc{c}}$ and becomes negative in the UV \cite{Grable2023}; like the scalar $\vec{\phi}^4$ and $\textrm{O}(N)^3$ theories, this fermionic theory is also asymptotically free at high energies. These models are interesting since calculations can be done analytically in the non-perturbative regime, and they realize non-trivial features like asymptotic freedom and phase transitions, and (at least in the case of the $\vec{\phi}^4$ model) bound states \cite{Romatschke:2022jqg,Romatschke:2023sce}.

Negative couplings, as long known by those studying $\PT$-symmetric non-Hermitian theories, can be meaningful and useful in a quantum field theory. The theory simply has to be path-integrated on some complex domain; the fields (e.g. $\vec{\phi}$) can no longer be real. However, $\PT$ (or some antilinear) symmetry of the Lagrangian and of the asymptotic Stokes wedges in which the domain terminates appears to be important for ensuring a physical result for quantities like the partition function. The existence of $\PT$-symmetric theories with complex or negative couplings, like the $\PT$-symmetric $-g\phi^4$ theory, indicates that a coupling constant need not be positive in order to have a predictive interacting theory. Although this is already known by many, I hope to emphasize it here to the broader community.

The implication of this present work, in line with prior work, is that the coupling constant of the $\vec{\phi}^4$ theory at large $N$ is only positive in the IR, while the bare coupling is negative. A negative bare coupling from the renormalization was already discussed in the 70s and 80s by Kobayashi and Kugo \cite{Kobayashi:1975ev}, by Bardeen and Moshe \cite{Bardeen:1983}, and by Stevenson, All\`es, and Tarrach \cite{Stevenson:1985,Stevenson:1987}, but the theory explicitly at negative coupling was never studied.  As mentioned before, a similar feature appears in the Lee model, which has a purely imaginary bare coupling \cite{Kleefeld:2004jb,Bender:2004sv}. Moreover, it has been pointed out that under Wilsonian renormalization, the effective action of a non-Hermitian $\PT$-symmetric theory can emerge from a Hermitian theory \cite{Bender:2021fxa}. Thus a sign change of the coupling under renormalization is expected to be a feature of other theories, too.

Life simplifies at large $N$ and it may be true that the proper $\PT$-symmetric $-g\vec{\phi}^4/N$ theory at large $N$, for example defined on the domain $\calC_{\PT}$ in \eqref{eq:domain-PT}, does in a sense extend the positive-coupling (Hermitian) theory above the scale $\LambdaLP$, as was first suggested by Romatschke in \cite{Romatschke:2022jqg}. My use here of the domain $\calC_-$ in \eqref{eq:domain-neg-coup} is motivated by the domain $\calC_{\PT}$ in \eqref{eq:domain-PT} on which the path integral of the $\PT$-symmetric theory could be defined. 

Moreover, although the domain $\calC_-$ considered in this paper does not correspond to a Hamiltonian theory with a $\calC\PT$ inner product of states, it still yields a partition function that agrees with that of the physical positive-coupling theory, and the theory on the domain $\calC_-$ may still be predictive beyond thermodynamic quantities like the pressure $p(\beta)$.

% One might object to the presence of asymptotic freedom in large-$N$ scalar and fermionic theories and in $\PT$-symmetric theories. Coleman and Gross in the 70s argued that the only asymptotically free theories are non-Abelian gauge theories. However, their argument assumed that scalar field theory with a negative coupling constant led to an unbounded potential \cite{PhysRevLett.31.851}. It is no longer true that the potential is unbounded from below if the scalar fields live on an appropriate complex domain, such as $\calC_{\PT}$ in \eqref{eq:domain-PT} or $\calC_-$ in \eqref{eq:domain-neg-coup}, or the $\PT$-symmetric domain considered in \cite{Bender:2006wt,Jones:2006qs} for the $-g\phi^4$ theory. 
% Thus scalar field theories can in fact be asymptotically free.

% An objection can also be raised specifically about the $\vec{\phi}^4$ model at large $N$. In the 70s, Coleman, Jackiw, and Politzer studied the $\textrm{O}(N)$ model at large $N$ in 3+1D and found it possesses a tachyon, and on this basis they argued that the theory is ``sick'' \cite{PhysRevD.10.2491}. However, it must be pointed out that this tachyon appears only in the dispreferred vacuum in which the field $\vec{\phi}$ has no mass gap. This instability goes away when one realizes that the correct vacuum is the one where the field has a mass gap $m=\sqrt{e}\LambdaLP$, as pointed out by Abbott, Kang, and Schnitzer \cite{PhysRevD.13.2212}. 

Here I have non-perturbatively calculated the partition function of the $\vec{\phi}^4$ theory at large $N$ assuming negative coupling from the start, with a careful choice of domain of path integration. This goes beyond prior work \cite{Ogilvie:2008tu}, but obtains a result consistent with that work. With the proper choice for the complex domain $\calC_-$ of path integration, involving $\PT$ symmetry, the resulting thermodynamics are the same as those in the positive-coupling $\textrm{O}(N)$ theory. The form of the running coupling in both theories indicates that the coupling, in fact, changes sign under renormalization, and the equivalence of thermodynamic observables under this change of sign suggests, in the spirit of renormalization, that the negative-coupling theory simply extends the positive-coupling theory to scales above that of the Landau pole.  Because of this, and because of other examples in large-$N$ and $\PT$-symmetric field theory, I point out that a negative bare coupling can be sensible. In addition, if the positive-coupling theory does flow under renormalization to the negative-coupling theory in the UV, as this work seems to hint, then the $\vec{\phi}^4$ theory at large $N$ would give an example of a scalar field theory with asymptotic freedom and which is not quantum trivial.

A next step would be to see if thermodynamics in the $\vec{\phi}^4$ theory (and the issue of $\PT$ symmetry breaking) is modified by the dominance of pressures associated with non-constant saddle configurations $m^2(x)$ of the auxiliary field. It would also be fruitful to calculate viscosities and other transport coefficients in the $\vec{\phi}^4$ theory in 3+1D along the lines of \cite{Romatschke:2021imm,Weiner:2022kgx,Lawrence:2022vwa}. One way to do this involves the use of R$n$ resummation methods \cite{Romatschke:2019wxc,Romatschke:2019rjk}. Lastly, it would be interesting to see if the results of this work continue at sub-leading order in the large-$N$ expansion.